\documentclass[12pt]{article}
\usepackage[xdvi]{graphicx} 
\usepackage{psfrag}
\usepackage{subfigure}
\usepackage{fancyhdr}
\usepackage{epsfig}
\usepackage{cite}
\usepackage{amsmath}
\usepackage{theorem}
\usepackage{amssymb}
\usepackage{latexsym}
\usepackage{epic}
\begin{document}  
\newcommand{\todo}[1]{{\em \small {#1}}\marginpar{$\Longleftarrow$}}   
\newcommand{\w}{\widehat}
\rightline{OUTP-03-08P}  
\rightline{gr-qc/0303082}
\vskip 1cm 
\begin{center}
{\Large \bf Surface terms and the Gauss-Bonnet Hamiltonian}
\end{center}
\vskip 1cm
  
\renewcommand{\thefootnote}{\fnsymbol{footnote}}   \centerline{\bf Antonio
Padilla\footnote{a.padilla1@physics.ox.ac.uk}}
\vskip .5cm   \centerline{ \it Theoretical Physics, Department
of  Physics,}   \centerline{\it University of Oxford, 1 Keble Road, Oxford OX1 3NP, U.K.}
  
\setcounter{footnote}{0} \renewcommand{\thefootnote}{\arabic{footnote}}
 

\begin{abstract}
We derive the gravitational Hamiltonian starting from the Gauss-Bonnet
action, keeping track of all surface terms. This is done using the
language of orthonormal frames and forms to keep things as tidy as
possible. The surface terms in the Hamiltonian give a remarkably simple expression for
the total energy of a spacetime. This expression is consistent with
energy expressions found in hep-th/0212292. However, we can apply our
results whatever the choice of background and whatever the symmetries
of the spacetime.
\end{abstract}

\newpage

\section{Introduction}
The energy of gravitational systems has attracted much interest down
the years (see, for
example~\cite{Arnowitt:ADMmass,Abbott:stability,
Brown:energy1,Brown:energy2, Hawking:hamiltonian, Brown:energy3, Brown:energy4, Brown:energy5,
Hayward:quasi-energy, Hayward:sph-energy, Pinto-Neto:energy, Bose:energy, Hayward:energy, Deser:quadenergy,Deser:HDenergy}).  In
particular, for Einstein gravity
\begin{equation} \label{EHaction}
S=\kappa \int R-2 \Lambda,  
\end{equation}
expressions were found for the energy of asymptotically flat~\cite{Arnowitt:ADMmass}
and asymptotically AdS~\cite{Abbott:stability} spacetimes. Some time
later, Hawking and
Horowitz~\cite{Hawking:hamiltonian} gave a general derivation of the gravitational
Hamiltonian, keeping careful track of surface terms. When
evaluated on a solution, this Hamiltonian gave an expression for the
total energy. This agreed with the earlier expressions found
in~\cite{Arnowitt:ADMmass,Abbott:stability}. However, this new
expression could be applied regardless of the asymptotic behaviour of
the solution.  

Recently, Deser and Tekin have found energy expressions for 
higher derivative
gravities~\cite{Deser:quadenergy,Deser:HDenergy}. This inital work has
focused on test spacetimes
that are {\it asymptotically} maximally symmetric, with background
spacetimes (vacua), 
defined to be maximally symmetric {\it everywhere}.  In this paper, we will
perform the analogue of Hawking and Horowitz's calculation by deriving
the Hamiltonian for Gauss-Bonnet (GB) gravity: a special, but important
class of higher derivative gravities. We will obtain an expression for
the energy by once again evaluating this Hamiltonian on a solution. As
with~\cite{Hawking:hamiltonian}, our expression will be consistent with earlier results~\cite{Deser:quadenergy,
Deser:HDenergy}, but can be easily applied even if the background is
not maximally symmetric. This greater flexibility allows for a more
natural choice of vacua in some cases, as we will illustrate with an example.

Before going any further, let us briefly mention what GB gravity is,
and why it is important. GB gravity is given by the addition of the
Gauss-Bonnet invariant to the Einstein-Hilbert action~(\ref{EHaction}),
\begin{equation} \label{GBaction1}
S= \kappa \int R-2 \Lambda +\alpha
\left(R^2-4R_{AB}R^{AB}+R_{ABCD}R^{ABCD} \right).
\end{equation}
In $D=4$ dimensions, the GB term  is a topological invariant and does
not enter the dynamics. This ceases to be the case in $D>4$
dimensions. Furthermore, in Einstein gravity, the vacuum field equation
are given by a linear combination of the Einstein tensor and the
metric. In four dimensions, this is the most general combination of
tensors that satisfies the following conditions~\cite{Lovelock:einstein}:
\begin{itemize}
\item it is symmetric.
\item it depends only on the metric and its first two derivatives.
\item it has vanishing divergence.
\item it is linear in the second derivatives of the metric\footnote{In
$D=4$, this condition is actually implied by the other three.}.
\end{itemize}
However, it has become unfashionable to restrict ourselves to just
four
dimensions. If we go to $D=5$ or $6$, it turns out
that these conditions are satisfied by a linear combination of the
metric, the Einstein tensor, and the {\it Lovelock
tensor}~\cite{Lovelock:einstein,Lanczos}. The Lovelock tensor arises
from the variation of the Gauss-Bonnet term in the above
action~(\ref{GBaction1}). As $D$ increases, we can use the same
arguments to introduce higher order invariants (the Euler
characters)~\cite{Myers:HDsurface}. In this sense, GB gravity is the natural
generalisation of Einstein gravity to $5$ or $6$
dimensions, although
other higher derivative gravity theories have also been studied (see,
for example~\cite{Hehl:quad, Kijowski:nonlin,
Kerner:inflation,Mueller:spon, Mueller:grav}). 

It is interesting to note that the GB action (\ref{GBaction1}) is
renormalisable~\cite{Stelle:ren}, but perhaps the most compelling reason to study GB gravity is its
appearance in String theory. Consider the slope (ie.~$\alpha'$)
expansion  for
the heterotic string. At lowest order, it is given by the
Einstein-Hilbert action~(\ref{EHaction}). The first stringy
corrections give rise to quadratic curvature terms~\cite{Candelas:vacconf,Green:D-inst}. For
this effective theory to be ghost-free, the quadratic curvatures must be in
the combination given in the GB
action~(\ref{GBaction1})~\cite{Zwiebach:curvsq,Zumino:gravth}. This
link to String theory has generated a lot of research into higher
derivative gravity and cosmology~\cite{Madore:KK, Madore:sing,
Deruelle:KKcosmo, Madore:cosmo, Kerner:cosmo,  Charmousis:gauss-bonnet, Binetruy:avoid,
Davis:israel, Gravanis:israel, Germani:inspired, Germani:vary, Nojiri:FRW,
Lidsey:bwcosmo, Nojiri:cosmo, Nojiri:bwcosmo,
Meissner:local, Neupane:consist, Neupane:potential, Neupane:local,
Cho:warped, Cho:thermal, Cho:noghost}.  Although the early work~\cite{Madore:KK, Madore:sing,
Deruelle:KKcosmo, Madore:cosmo, Kerner:cosmo} achieved
reduction to four dimensions {\it via} Kaluza-Klein compactifications,
more recent studies~\cite{ Charmousis:gauss-bonnet, Binetruy:avoid,
Davis:israel, Gravanis:israel, Germani:inspired, Germani:vary, Nojiri:FRW,
Lidsey:bwcosmo, Nojiri:cosmo, Nojiri:bwcosmo,
Meissner:local, Neupane:consist, Neupane:potential, Neupane:local,
Cho:warped, Cho:thermal, Cho:noghost} have focussed on the braneworld scenario~\cite{Randall:hierarchy,Randall:compactification}. Braneworld models are themselves inspired by
String theory~\cite{Horava:heterotic} so it is natural to ask what effect any
stringy corrections might have on their cosmology.  From a holographic
point of view, we might expect such higher curvature terms in the bulk
to correspond to next to leading order corrections in the $1/N$
expansion of the CFT on the boundary/brane~\cite{Nojiri:bwcosmo}. Calculating the GB
Hamiltonian will allow us to investigate the GB
version of ``exact'' braneworld holography~\cite{Padilla:exact, Padilla:thesis}. This will be discussed in a future article~\cite{Padilla:GBholog}.

The rest of this paper is organised as follows: in section~2 we will
give a derivation of the GB Hamiltonian, with extra details given in
the appendix. We will use orthonormal
frames and differential forms to keep things as tidy as possible. Some
readers may wish to ignore the details of this derivation and go
directly to the energy expression at the end of the section. In
section~3 we will show that this expression is consistent with existing
literature~\cite{Crisostomo:BHscan,Deser:quadenergy,Deser:HDenergy}.
We will illustrate the flexibility of our approach with a special
example in section~\ref{specialcase}. Section~4 contains some
concluding remarks. 

\section{Derivation of the Gauss-Bonnet Hamiltonian}
\subsection{The action}
The Gauss-Bonnet action~(\ref{GBaction1}) is most elegantly written in
terms of differential forms. Suppose our $D$-dimensional spacetime,
$\mathcal{M}$, has metric,
\begin{equation}
g=\eta_{AB}E^A \otimes E^B
\end{equation}
where $\{E^A \}$ is an orthonormal basis of 1-forms, and indices are
raised/lowered using
$\eta_{AB}=\textrm{diag}(-1, +1, \ldots, +1)$. We write $\{ X_A \}$
for the
dual basis of vectors.

We will find it useful
to define the following forms,
\begin{equation}
e_{A_1 \ldots A_m}=\frac{1}{(D-m)!}\epsilon_{A_1 \ldots A_m A_{m+1}
\ldots A_D} E^{A_{m+1}} \wedge \ldots \wedge E^{A_D}, \label{measure1}
\end{equation}
where $\epsilon_{A_1 \ldots A_D}$ is the totally antisymmetric tensor
with $\epsilon_{0 \ldots (D-1)}=1$. Notice that the scalar $D$-form, $e$, is the volume
measure on $\mathcal{M}$.

Now suppose we have vanishing torsion, and that $\omega^A{}_B$ is the
connection $1$-form compatible with the metric $g$. The curvature $2$-form is given by 
\begin{equation} \label{2form}
\Omega^A{}_B=d\omega^{A}{}_{B}+\omega^{A}{}_{C} \wedge
\omega^C{}_{B}=\frac{1}{2}R^A{}_{BCD} E^C \wedge E^D.
\end{equation}
The right hand equation above gives the Riemann tensor, $R^A{}_{BCD}$. The
Ricci tensor is then defined by $R_{BD}=R^C{}_{BCD}$  and the Ricci
scalar by $R=\eta^{AB}R_{AB}$. The GB action~(\ref{GBaction1}) can now
be written
\begin{equation} \label{GBaction2}
S=\kappa \int_{\mathcal{M}} -2\Lambda e + \Omega^{AB} \wedge e_{AB}
+\alpha  \Omega^{AB} \wedge \Omega^{CD}  \wedge e_{ABCD}.
\end{equation}
where we have made use of the following
identity~\cite{Myers:HDsurface, Mueller:spon},
\begin{equation}
E^B \wedge e_{A_1 \ldots A_m}=\delta^B_{A_m} e_{A_1 \ldots A_{m-1}}-
\delta^B_{A_{m-1}} e_{A_1 \ldots
A_{m-2}A_m}+\ldots+(-1)^{m-1}\delta^B_{A_1} e_{A_2 \ldots A_m}, \label{useful}
\end{equation}

If $\mathcal{M}$ has a boundary, $\partial \mathcal{M}$, we need
to define boundary conditions on $\partial \mathcal{M}$. We usually
demand that the geometry of the boundary is fixed. However, as it
stands, the action (\ref{GBaction2}) is inconsistent with these
boundary conditions. This is because its variation with respect to the
metric does not vanish on
shell.

To cure this, we need to add a
boundary integral. In Einstein gravity, this is the well known
Gibbons-Hawking term~\cite{Gibbons:GHterm}. The generalisation to
higher derivative gravities was worked out by
Myers~\cite{Myers:HDsurface}. For GB gravity it is given by
\begin{equation}
S_{\textrm{boundary}}=-\kappa \int_{\partial \mathcal{M}} \theta^{AB}
\wedge e_{AB} + 2\alpha  \theta^{AB}\wedge \left
( \Omega^{CD}-\frac{2}{3} \theta^C{}_E \wedge \theta^{ED}\right)\wedge e_{ABCD}, \label{bdy}
\end{equation}
where $\theta^{AB}$ is the second fundamental
form~\cite{Eguchi:diffgeom, Choquet:analysis}. It is defined as
\begin{equation}
 \theta^{AB} = \omega^{AB}-\omega^{AB}_0.
\end{equation}
where $\omega^{AB}_0$ is the connection for the {\it product}
metric, $g_0$, that agrees with $g$ on the boundary. The second fundamental form is closely related to the
extrinsic curvature of $\partial \mathcal{M}$ in
$\mathcal{M}$~\cite{Myers:HDsurface}. 

The fully consistent action for Gauss-Bonnet gravity
is therefore given by,
\begin{equation} \label{b+baction}
S=S_{\textrm{bulk}}+S_{\textrm{boundary}}
\end{equation}
where $S_{\textrm{bulk}}$ and $S_{\textrm{boundary}}$ are given by
equations~(\ref{GBaction2}) and (\ref{bdy}) respectively. It is
interesting to note that the boundary integrand contains a linear
combination of the first two Chern-Simons forms. This is related to
the fact that the bulk integrand contains the same linear combination
of the first two Euler characters~\cite{Myers:HDsurface}.

Finally, we note that although the action (\ref{b+baction}) is well
defined for spatially compact manifolds, it is divergent when
$\mathcal{M}$ is spatially noncompact~\cite{Hawking:hamiltonian}. To get round
this we need to choose a reference background, $\mathcal{\bar M}$ with
metric $\bar g$. This background should be a static solution to the
field equations~\cite{Hawking:hamiltonian}, but does not have to be
maximally symmetric. The boundary conditions are unchanged
which means that  $\partial \mathcal{ \bar M}$ should have the same
geometry as  $\partial \mathcal{M}$. We can then define the background
action, $\bar S$, in the same way as before. The physical action is the
difference,
\begin{equation} \label{physact}
I=\Delta S=S-\bar S.
\end{equation} 
In the Hamiltonian picture, this background can be thought of as
defining a background energy or zero energy solution. For example, for
an asymptotically AdS spacetime, we would probably choose the
background to be pure AdS, but we do not have to. Any asymptotically
AdS black hole spacetime would be equally valid.
\subsection{Splitting space and time}
Although the physical Hamiltonian will be constructed from the
action~(\ref{physact}), it is clear that it will just be the difference
of the Hamiltonian constructed from $S$ and that constructed from
$\bar S$~\cite{Hawking:hamiltonian}. For the time being we will concentrate on the former.
\subsubsection{Foliations of $\mathcal{M}$} \label{Mfoliations}
To proceed, we need to deconstruct the spacetime $\mathcal{M}$ by
separating space from time in the following way. First, we choose a
timelike vector field, $\partial/\partial t$. Now introduce a
family of spacelike hypersurfaces $\{\Sigma_t \}$ labelled by
the parameter $t$. This family is a
foliation of the full spacetime. We assume that the hypersurfaces have no
inner boundaries and do not intersect each other. They
meet the  timelike part of the boundary (call this $B$)
orthogonally, and in the far past/future, they coincide with
the spacelike part of the boundary (call this
$\Sigma_{\infty}$). Therefore the
total boundary, $\partial \mathcal{M}=B \cup \Sigma_{\infty}$.

We can write the metric for
$\mathcal{M}$ in ADM form~\cite{Arnowitt:ADMmass},
\begin{equation} \label{ADM}
g=-N^2dt^2+\gamma_{ab}(t, x^a)(dx^a+N^adt)(dx^b+N^bdt),
\end{equation}
where $N$ is the lapse function, $N^a$ the shift vector, and
$\gamma_{ab}(t, x^a)$ the induced metric on $\Sigma_t$. It is natural
to choose the following orthonormal basis of $1$-forms,
\begin{equation}
E^\bot=Ndt, \qquad E^a=E^a{}_b(dx^b+N^bdt)
\end{equation}
where $\delta_{ab}E^a{}_c E^b{}_d=\gamma_{cd}$. We would like to emphasize some notation at this point. Lowercase latin indices
label components in  $\Sigma_t$,  whereas
uppercase latin indices label components in $\mathcal{M}$. For
example,  $\{ E^a \}$ is an orthonormal basis for
$\Sigma_t$ whereas $\{ E^A \}=E^\bot \cup \{ E^a \}$ is an orthonormal
basis for  $\mathcal{M}$.

The dual basis of vectors is given by~\cite{Lau:diffforms}
\begin{equation} \label{dualbasis}
X_\bot=\frac{1}{N} \left( \frac{\partial}{\partial t}
-N^a\frac{\partial}{\partial x^a} \right), \qquad X_a=E_a{}^b\frac{\partial}{\partial x^a}.
\end{equation}
$X_\bot$ is the vector normal to  $\Sigma_t$, and is not
necessarily equal, or even parallel to $\partial/
\partial t$. We should also note that  $X_\bot$ is
tangent to $B$, as $\Sigma_t$ and $B$ are orthogonal. 

Normally, the next step is to use the Gauss-Codazzi
equations~\cite{Wald:Gauss-Codacci} to rewrite the bulk part of the action (see for
example~\cite{Hawking:hamiltonian}). This has
infact been done for GB gravity~\cite{Teitelboim:GBham} although the contribution
from surface terms was ignored. In this paper we are using the
language of orthonormal frames and differential forms. We therefore
need to know how to translate the Gauss-Codazzi equations into this
language. This is explained in~\cite{Isenberg:canonical}, so we will
merely state their results.

The Gauss-Codazzi equations describe the decomposition of the bulk
Riemann tensor into spatial tensors defined on $\Sigma_t$. In the language of forms, it is the
curvature $2$-form that we wish to decompose. We start by decomposing
the connection, $\omega^{AB}$.
\begin{eqnarray}
\omega^{\bot a} &=& -H^a+a^a E^\bot \label{con1}, \\
\omega^{ab} &=& \quad\tilde \omega^{ab}+l^{ab} E^\bot . \label{con2}
\end{eqnarray}
Here, we have two $0$-forms: a vector,  $a^a$, and an 
antisymmetric tensor, $l^{ab}$.  The $1$-form $H^a=H^a{}_bE^b$, where $H^a{}_b$ is the extrinsic curvature of $\Sigma_t$ in
$\mathcal{M}$. From now on, anything labelled with a tilde is
intrinsic to $\Sigma_t$, as opposed to $\mathcal{M}$. Therefore, $\tilde \omega^{ab}$ is the connection for the
{\it induced} metric  $\gamma=\delta_{ab}E^a \otimes E^b$. 

We now write the Gauss-Codazzi equations in the following way
\begin{eqnarray} 
\Omega^{\bot a} &=&-\tilde \nabla H^a + E^\bot \wedge \left[-\$_\bot H^a-\frac{1}{N} \tilde \nabla(Na^a)-l^{ab}H_b \right]  \label{GC1}\\
\Omega^{ab} &=& \tilde \Omega^{ab}+H^a \wedge H^b+E^\bot \wedge \left[ \$_\bot \tilde \omega^{ab}-\frac{1}{N} \tilde \nabla(Nl^{ab})+H^a
a^b-H^ba^a \right] \label{GC2}
\end{eqnarray}
where $\tilde \Omega^{ab} =\tilde d \tilde \omega^{ab} +\tilde
\omega^{a}{}_c \wedge \tilde \omega^{cb}$ is the curvature $2$-form
for $\Sigma_t$, and the operator $\tilde \nabla$ is the {\it covariant exterior
derivative}~\cite{Eguchi:diffgeom, Choquet:analysis}  (on $\Sigma_t$). Note that $\tilde \nabla$ reduces to the
exterior derivative, $\tilde d$, when acting on scalars. The definition of the operator $\$_\bot$
is given in~\cite{Isenberg:canonical}. It is closely related to the
Lie derivative with respect to the vector $X_\bot$, although it lives
entirely on $\Sigma_t$ and acts on tensor components as if they were
scalars. In many ways it behaves like a partial derivative\footnote{It
is often useful to think of $\$_\frac{\partial}{\partial t}$ as the
frame-form version of $\frac{\partial}{\partial t}$.}.

In principle we could also decompose the torsion $2$-form (see~\cite{Isenberg:canonical}). However, we
have set torsion to zero, which means that every component of the torsion
decomposition must vanish. This  gives the following conditions:
\begin{eqnarray} 
H^{ab} &=& H^{ba}, \label{torsion1} \\
a_a E^a &=& \frac{\tilde d N}{N},  \label{torsion2}\\
\tilde \nabla E^a &=& 0  \label{torsion4} \\
\$_\bot E^a &=& -H^a -l^a{}_b E^b. \label{torsion3}
\end{eqnarray}

\subsubsection{Foliations of $B$} 
Since we intend to keep careful track of surface terms, we will need a
foliation of $B$, as well as
$\mathcal{M}$. On $B$, the foliation is given by the family of
surfaces $\{ S_t \}$.  For a given value of $t$, $S_t$ is the
intersection of $B$ and $\Sigma_t$.

We need to understand how bulk quantities project on to $S_t$. Near
$S_t$, the metric can be written in Gaussian normal coordinates
\begin{equation} \label{GNnearB}
g=dz^2-E^\bot \otimes E^\bot +\delta_{ij}E^i \otimes E^j
\end{equation}
From now on we will write $E^z$ for the
extension of $dz$ into the bulk, and $X_z$ for the extension of the
inward pointing 
normal $\partial/\partial z$. Notice that we are using indices $i$, $j$ etc to label components in $S_t$.  

We can use the techniques developed in~\cite{Isenberg:canonical} to
project bulk quantities on to $B$ and then on to $S_t$. We find that
the decomposition of the connection is given by
\begin{align}
&\mspace{165mu}& \omega^{\bot z}&=& &b_i E^i& &+&
&cE^\bot& &&&&&\qquad\qquad  \label{con5}\\
&\mspace{165mu}& \omega^{\bot i}&=& -&\w H^{i}& &+&  &a^i E^\bot&
&+& &b^i E^z& &\qquad\qquad \label{con6}\\
&\mspace{165mu}&\omega^{zi}&=& &\w K^{i}& &+& &b^iE^\bot& &&&&  &\qquad\qquad \label{con7}\\           \
&\mspace{165mu}&\omega^{ij}&=&  &\w \omega^{ij}& &+& &l^{ij}E^\bot&
&+& &\chi^{ij} E^z& &\qquad\qquad  \label{con8}
\end{align}
where $\chi^{ij}$ is
some antisymmetric tensor, and
\begin{eqnarray}
c &=& a^z \label{c} \\
b^i &=& l^{zi}=-H^{zi} \label{b}
\end{eqnarray}
Note that anything wearing a hat
is intrinsic to $S_t$. Therefore, $\w \omega^{ij}$ is the
connection for the induced metric (on $S_t$), $\lambda=\delta_{ij}E^i
\otimes E^j$.  We have
also defined,
\begin{eqnarray}
\widehat H^i &=& H^i{}_jE^j,  \label{hatH}\\
\widehat K^i &=&K^i{}_jE^j. \label{hatK}
\end{eqnarray}
Here, we should  interpret $H^i{}_j$ and
$K^i{}_j$ as the extrinsic curvatures of $S_t$ in $B$ and $\Sigma_t$
respectively. 

The curvature decomposition is given by
\begin{eqnarray}
\Omega^{\bot z} &=&  \w d \left( b_i E^i \right)+ \w H^i \wedge \w
K_i+E^z\wedge\{\cdots\}\nonumber  \\
&& \qquad +E^\bot \wedge \left[ \w \$_\bot \left(b_i E^i \right)
-\frac{1}{N} \w d (Nc) -a^i \w K_i -b^i \w H_i \right]\label{tony0} \\
\Omega^{\bot i} &=& -\widehat \nabla \widehat H^i +b_j E^j \wedge
\widehat K^i +E^z\wedge\{\cdots\}\nonumber \\
&& \qquad +E^\bot \wedge \left[ - \widehat \$_\bot \widehat H^i
-\frac{1}{N}\widehat \nabla
(Na^i)-l^{ij} \widehat H_j +c\widehat K^i -b^i b_jE^j
\right] \label{tony1} \\
\Omega^{zi} &=& \widehat \nabla \widehat K^i-b_jE^j\wedge
\widehat H^i +E^z\wedge\{\cdots\}\nonumber \\
&& \qquad
+E^\bot \wedge \left[ \widehat \$_\bot \widehat K^i
-\frac{1}{N}\widehat \nabla (Nb^i)+\widehat K_j l^{ij} -b_jE^j
a^i-c \widehat H^i \right] \label{tony2}\\
\Omega^{ij} &=& \widehat \Omega^{ij}+\widehat H^i \wedge \widehat H^j
-\widehat K^i \wedge \widehat K^j +E^z\wedge\{\cdots\}\nonumber \\
&&\qquad+E^\bot \wedge\left[\widehat \$_\bot
\widehat \omega^{ij}-\frac{1}{N}\widehat \nabla (Nl^{ij})+2\widehat
H^{[i}a^{j]}+2\widehat
K^{[i} b^{j]} \right]\label{tony3}
\end{eqnarray}
where we will not care what is contained in $\{ \cdots \}$ as the
integration of $E^z$ over $B$ is zero. In analogy with the previous
section, $\w \Omega^{ij}$ and $\w \nabla$ are the curvature form and
covariant exterior derivative on $S_t$, respectively. Again, $\w \nabla$ reduces to
the exterior derivative, $\w d$, when acting on scalars. The operator
$\w \$_\bot$ is the analogue of $ \$_\bot$ on $B$~\footnote{If $\tilde
A$ is an arbitrary $p$-form in $\Sigma_t$, then near $S_t$,
we can write $\tilde A=\widehat A +A_z \wedge E^z$, where $\widehat A$ and $A_z$ are $p$ and $(p-1)$-forms respectively,
living on $S_t$. It can then be shown that  
$\$_\bot \tilde A =\w \$_\bot \w A +E^z \wedge\{\cdots\}$.}.

As long as we are near $B$, equations (\ref{tony0}) to (\ref{tony3}) are  the frame-form
version of the Gauss-Codazzi equations for a hypersurface of
codimension two~\cite{Eisenhart:GaussCod}.

\subsection{The Hamiltonian}
We are now ready to start calculating the Hamiltonian. However, we
will find it convenient to continue working with the action until
virtually the bitter end. When our action finally has the desired form we will switch to the
Hamiltonian picture, and give an expression for the gravitational
energy of a solution.

We will start with the bulk part of the action (\ref{GBaction2}). Our
aim is to write it so that it contains no derivatives of the lapse
function or the shift vector. This is because these are ignorable
coordinates, and should behave like Lagrange multipliers. They will
be paired with the Hamiltonian and momentum constraints respectively,
as is the case in Einstein gravity~\cite{Hawking:hamiltonian}. We also want to eliminate terms like $\$_\bot H^a$, which
contain second time derivatives of the canonical variable $E^a$.  We
will need to use integration by parts to achieve these aims. This
means that the bulk action (\ref{GBaction2}) will contribute surface
terms. In summary, we expect to write (\ref{GBaction2}) as
\begin{equation} \label{formofbulk}
S_\textrm{bulk}=S^*_\textrm{bulk}+S_\textrm{leftover},
\end{equation}
where $S_\textrm{leftover}$ are the leftover surface terms, and 
\begin{equation} \label{*bulk1}
S^*_\textrm{bulk} = \kappa \int dt \int_{\Sigma_t} \pi_a \wedge
\dot E^a -N\mathcal{H}-N^a
\mathcal{H}_a.
\end{equation}
Here $\pi_a$ is the momentum conjugate to $E^a$, and $\mathcal{H}$
and $\mathcal{H}_a$ are the Hamiltonian and momentum constraints
respectively. We have also introduced the intuitive notation  $\dot  E^a =\$_\frac{\partial}{\partial t}E^a$. 

At this stage, the surface part of the action is given
by equation~(\ref{bdy}). We can split this into two parts,
\begin{equation} \label{bdysplit}
S_{\textrm{boundary}}=S_B+S_\infty,
\end{equation}
where $S_B$ contains the integration over $B$ and $S_\infty$ the
integration over $\Sigma_\infty$. In section~\ref{hambdysec}, we will
group these terms with  $S_\textrm{leftover}$. This will give us a
modified boundary term
\begin{equation}
S^*_\textrm{boundary}=S_\textrm{boundary}+S_\textrm{leftover}.
\end{equation}
which will be closely related to the gravitational energy of a solution. 
\subsection{The bulk}
As promised, we start with the bulk part of the action (\ref{GBaction2}), in the hope of
deriving $S^*_\textrm{bulk}$ and $S_\textrm{leftover}$. Making use of
the formula (\ref{useful}), we find that the  bulk action
(\ref{GBaction2}) is given by
\begin{equation}
S_\textrm{bulk}=S_\textrm{kinetic}+S_1+S_2+S_3,
\end{equation}
where
\begin{eqnarray}
S_\textrm{kinetic} &=& \kappa \int_\mathcal{M} E^\bot \wedge \biggl\{ -2 \$_\bot H^b
\wedge \zeta_b  \biggr. \nonumber \\
&& \biggl. \qquad\qquad\qquad -4\alpha \left[ \tilde \nabla H^b \wedge \$_\bot
\tilde \omega^{cd} +\$_\bot H^b \wedge F^{cd} \right]\wedge \zeta_{bcd}
\biggr\} \label{skin}\\
S_1 &=& \kappa \int_{\mathcal{M}} E^\bot\wedge \biggl\{ -2\Lambda
\zeta +F^{ab} \wedge \zeta_{ab}+\alpha F^{ab} \wedge F^{cd} \wedge
\zeta_{abcd} \biggr\}\\
S_2 &=& \kappa \int_{\mathcal{M}} E^\bot\wedge \biggl\{ -\frac{2}{N}
\tilde \nabla (N a^b) \wedge \zeta_b  \biggr. \nonumber \\
&& \biggl. \qquad\qquad\qquad -4\alpha \left[\frac{1}{N} \tilde \nabla (N
a^b)\wedge F^{cd} +2 \tilde \nabla H^b \wedge H^{[c} a^{d]} \right] \wedge
\zeta_{bcd} \biggr\} \\
S_3 &=& \kappa \int_{\mathcal{M}} E^\bot\wedge \biggl\{4\alpha \left
[ \tilde \nabla H^b \wedge \frac{1}{N} \tilde \nabla (N \tilde
l^{cd}) -l^{ba}H_a \wedge F^{cd} \right] \wedge \zeta_{bcd} \biggr\}. \label{S3}
\end{eqnarray}
with
\begin{equation}
F^{ab}= \tilde
\Omega^{ab} + H^a \wedge H^b 
\end{equation}
Note that we have introduced the analogue of $e_{A_1 \ldots A_m}$ on
$\Sigma_t$,
\begin{equation}
\zeta_{a_1 \ldots a_m}=e_{\bot a_1 \ldots
a_m}=\frac{1}{(D-1-m)!}\epsilon_{a_1 \ldots a_m a_{m+1} \ldots
a_{D-1}}E^{a_1} \wedge \ldots \wedge E^{a_{D-1}}. \label{measure2}
\end{equation}
We will now rewrite expressions (\ref{skin}) to (\ref{S3}), bearing
in mind the goals we mentioned at the beginning of this section.
The kinetic term is given by
\begin{eqnarray}\label{kinetic3}
S_\textrm{kinetic} &=&\kappa \int dt \int_{\Sigma_t} \pi_a \wedge
\dot E^a -N^a \mathcal{H}_a \nonumber \\
&&+\kappa \int dt \int_{S_t} 4 \alpha NH^b \wedge \$_\bot
\tilde \omega^{cd}\wedge \zeta_{bcd}+(-1)^D N^a \pi_b E^b{}_a -S_\infty. 
\end{eqnarray}
where
\begin{eqnarray}
\pi_a &=& -2\zeta_{ab} \wedge H^b -4\alpha \zeta_{abcd} \wedge H^b \wedge
\Lambda^{cd}  \label{conjmom} \\
\Lambda^{cd} &=& \tilde \Omega^{cd} +\frac{1}{3} H^c \wedge H^d \label{lcd} 
\end{eqnarray}
and the Hamiltonian constraint, 
\begin{equation} \label{momentum}
\mathcal{H}_a=(-1)^{(D-1)} E^b{}_a \tilde \nabla \pi_b.
\end{equation}
In section~\ref{sec:conjmom}, we will show that $\pi_a$ is indeed the
momentum conjugate to $E^a$. The derivation of equation (\ref{kinetic3}) is given in
the appendix. 

The remaining terms in the bulk action can be written as follows:
\begin{eqnarray}
S_1 &=& -\kappa \int dt \int_{\Sigma_t} N\mathcal{H} \label{s1} \\
S_2 &=& \quad\kappa \int dt \int_{S_t} N\left[2a^b \zeta_b+4
\alpha a^b F^{cd}\wedge
\zeta_{bcd}\right] \label{s2}\\
S_3 &=&  -\kappa\int dt \int_{S_t} 4\alpha N \tilde \nabla H^b l^{cd} \wedge
\zeta_{bcd} \label{s3}
\end{eqnarray}
where the Hamiltonian  constraint, $\mathcal{H}$,  is given by
\begin{equation} \label{hamiltonian}
\mathcal{H}=2\Lambda \zeta-F^{ab} \wedge \zeta_{ab}-\alpha F^{ab}
\wedge F^{cd } \wedge\zeta_{abcd}.
\end{equation}
Note that in deriving the expression (\ref{s3}) for $S_3$, we made use
of the following:
\begin{equation}
\tilde \nabla^2 H^b= \tilde \Omega^b{}_c \wedge H^c.
\end{equation}
\begin{equation} \label{shankly}
H_a \wedge\left[\tilde
l^{ba}F^{cd} +l^{cd} \tilde \Omega^{ba} \right]\wedge
\zeta_{bcd}=0.
\end{equation}
Collecting together equations (\ref{kinetic3}), (\ref{s1}), (\ref{s2}),
and (\ref{s3}), we see that $S_\textrm{bulk}$ takes the desired form
\begin{equation}
S_\textrm{bulk}=S^*_\textrm{bulk}+S_\textrm{leftover}
\end{equation}
with $S^*_\textrm{bulk}$ given by equation
(\ref{*bulk1}) and $S_\textrm{leftover}$ given by
\begin{multline} \label{leftover}
S_\textrm{leftover} = -S_\infty+ \kappa \int dt \int_{S_t} (-1)^D
N^a \pi_b E^b{}_a  \\
+ N\left\{2a^b \zeta_b +4\alpha \left[ H^b \wedge \$_\bot
\tilde \omega^{cd}+ a^b F^{cd}
-\tilde \nabla H^b l^{cd}
\right]
\wedge
\zeta_{bcd}\right\}.
\end{multline}
\subsection{The boundary} \label{hambdysec}
As expected, rewriting the bulk part of the action in the desired
form (\ref{*bulk1}) has altered the boundary part of the action.  In
particular, we have a leftover surface integral (\ref{leftover}) that
must be added to the original boundary part of the action
(\ref{bdysplit}). This gives the modified boundary action,
\begin{equation} \label{*bdy}
S^*_\textrm{boundary}=S_\textrm{leftover}+S_\infty+S_B.
\end{equation}
In order to combine each term in the above equation, we need to write
them in a common form. This will involve integrations over $S_t$, of
well defined quantities on $S_t$. 

Let us begin with $S_\textrm{leftover}+S_\infty$. From (\ref{GNnearB})
we know that near $B$, $E^z{}_a=\delta^z{}_a$ and $N^z=0$. This means
that the momentum term in (\ref{leftover}) gives
\begin{equation} 
\kappa \int dt \int_{S_t} (-1)^D N^a \pi_b E^b{}_a=\kappa \int dt
\int_{S_t} (-1)^D N^i\pi_j E^j{}_i.
\end{equation}
We now use the fact that
\begin{equation}
\$_\bot \tilde \omega^{zi} =  \w \$_\bot \w K^i +E^z \{\cdots
\}. \label{dollarK}
\end{equation}
along with (\ref{c}) and (\ref{b}) to rewrite the remaining terms. We
find that
\begin{multline}
S_\textrm{leftover}+S_\infty= \kappa \int dt \int_{S_t} (-1)^DN^i
\pi_j E^j{}_i +2Nc\phi\\
+ 4 \alpha N\left[ -b_k E^k \wedge \w \$_\bot \w
\omega^{ij}+cF^{ij}+F^z l^{ij} -2\w H^i \wedge \w \$_\bot \w K^j -2a^i F^{zj}-2F^i b^j \right] \wedge
\phi_{ij} \label{Sleft+infinity}
\end{multline}
where $F^b=-\tilde \nabla H^b$. We have also introduced the $S_t$ analogue of $\zeta_{a_1
\cdots a_n}$ and $e_{A_1 \cdots A_n}$:
\begin{equation} \label{measure3}
\phi_{i_1 \cdots i_n}=\zeta_{z i_1 \cdots i_n}=e_{\bot z i_1 \cdots
i_n}.
\end{equation}
Meanwhile, terms like $F^{z}$ and $F^{ij}$ can be deduced by comparing
equations (\ref{tony0}) to (\ref{tony3}) with (\ref{GC1}) and (\ref{GC2}).
\begin{eqnarray}
F^z &=&  \w d \left( b_k E^k \right)+ \w H^k \wedge \w
K_k \label{Fz}\\ 
F^i &=& -\widehat \nabla \widehat H^i +b_k E^k \wedge
\widehat K^i  \label{Fi}\\
F^{zj} &=& \widehat \nabla \widehat K^j-b_kE^k\wedge
\widehat H^j \label{Fzj}\\
F^{ij} &=& \widehat \Omega^{ij}+\widehat H^i \wedge \widehat H^j
-\widehat K^i \wedge \widehat K^j \label{Fij}
\end{eqnarray}
Note that we have dropped all terms like $E^z \{\cdots \}$ as we are
now integrating over $B$.

Now consider $S_B$. The only non-zero components of $\theta^{AB}$ on
$B$ are:
\begin{equation}
\theta^{\bot z}=\omega^{\bot z}, \qquad \theta^{zi}=\omega^{zi}
\end{equation}
where $\omega^{\bot z}$ and $\omega^{zi}$ are given by equations
(\ref{con5}) and (\ref{con7}) respectively. We use this fact, along with equations (\ref{tony1}) and (\ref{tony3}),
to write $S_B$ in the following way:
\begin{eqnarray}
S_B &=& \kappa \int dt \int_{S_t}  2N \w K^i \wedge \phi_i -2Nc \phi
+4\alpha N\w K^i \wedge \left[ F^{jk} +\frac{2}{3} \w
K^j \wedge \w K^k \right] \wedge \phi_{ijk}\nonumber \\
&& +4\alpha N \left\{-2b^i \left[ F^j -\frac{2}{3} b_k E^k \wedge \w K^j\right]+2 \w K^i
\wedge \left[ G^j +\frac{2}{3} \left(-c \w K^j +b^j b_k
E^k\right)\right]
 \right.\nonumber\\
&&\left.\qquad\qquad -c\left[ F^{ij}+\frac{2}{3} \w K^i \wedge \w K^j
\right]+b_k E^k \wedge \left[ G^{ij} +\frac{4}{3}b^{[i} \w
K^{j]}\right] \right\} 
 \wedge \phi_{ij}\label{SB}
\end{eqnarray}
where
\begin{eqnarray}
G^{j} &=& - \widehat \$_\bot \widehat H^j
-\frac{1}{N}\widehat \nabla
(Na^j)-l^{jk} \widehat H_k +c\widehat K^j -b^j b_kE^k \\
G^{ij} &=& \widehat \$_\bot
\widehat \omega^{ij}-\frac{1}{N}\widehat \nabla (Nl^{ij})+2\widehat
H^{[i}a^{j]}+2\widehat
K^{[i} b^{j]} 
\end{eqnarray}
The expressions (\ref{Sleft+infinity}) and (\ref{SB}) now have a common
form, so we can combine them to get
\begin{equation}
S^*_\textrm{boundary}=S_{\textrm{boundary}, 1}+S_{\textrm{boundary}, 2}+S_{\textrm{boundary}, 3}
\end{equation}
where
\begin{eqnarray}
S_{\textrm{boundary}, 1} &=& \kappa \int dt \int_{S_t} (-1)^D N^i
\pi_j E^j{}_i 
+N \left\{ 2 \w K^i \wedge \phi_i \right. \nonumber\\
&&\left.\mspace{90mu} +4\alpha  \w K^i \wedge
\left[ \w \Omega^{jk} +\w H^j \wedge \w H^k -\frac{1}{3} \w K^j \wedge \w K^k \right] \wedge
\phi_{ijk} \right\} \label{*Sb1}\\
S_{\textrm{boundary}, 2} &=& \kappa \int dt \int_{S_t} 4 \alpha N
\left \{ \w d \left(b_k E^k\right) l^{ij}-\frac{1}{N} b_k E^k
\wedge \w
\nabla ( N l^{ij} ) \right. \nonumber \\
&& \left. \mspace{150mu} -2 a^i \w \nabla \w K^j -\frac{2}{N} \w K^i \wedge
\w \nabla ( Na^j ) \right\} \wedge \phi_{ij}  \label{Sb2}\\
S_{\textrm{boundary}, 3} &=& \kappa \int dt \int_{S_t}4 \alpha N
\left \{ -2 \w H^i \wedge \w \$_\bot \w K^j -2 \w K^i \wedge \w
\$_\bot \w H^j \right. \nonumber \\
&& \left. \mspace{150mu} +\w H^k \wedge \w K_k l^{ij} -2 \w K^i
\wedge l^{jk} \w H_k \right\}\wedge \phi_{ij}\label{Sb3}
\end{eqnarray}
Since we have zero torsion, we use the condition
\begin{equation}
\w \nabla  \phi_{i_1 \cdots i_n} \equiv 0
\end{equation}
to show that the integrand in $S_{\textrm{boundary}, 2}$ is a total
derivative. Since $S_t$ is a boundary of $\Sigma_t$, it has no boundary of its
own. This means that, 
\begin{equation} \label{*Sb2}
S_{\textrm{boundary}, 2}=0
\end{equation}
By integrating by parts on the time derivatives in
(\ref{Sb3}), we can also show that
\begin{equation} \label{*Sb3}
S_{\textrm{boundary}, 3}= \kappa \int dt \int_{S_t} 4 \alpha N \left\{ -2 \w K^i
\wedge \w H^j \wedge \w H^k \wedge \phi_{ijk} \right\}.
\end{equation}
Here we have used the zero torsion result
\begin{equation} \label{Sttorsion}
\w \$_\bot E^i = -\w H^i - l^i{}_j E^j.
\end{equation}
and the identity
\begin{multline} \label{ianrush}
\w H^k \wedge \w K_k l^{ij}\wedge \phi_{ij} -2\w K^i
\wedge \w H_k l^{jk}\wedge \phi_{ij}-2 \w K^i \wedge \w H^j
l^k{}_l \wedge E^l \wedge \phi_{ijk} \\
= \w K^i \wedge E_i \wedge \w H^j l^{kl} \wedge
 \phi_{jkl}=0. 
\end{multline}
The last equality in (\ref{ianrush}) follows from the symmetry of $\w K^{ij}$.

We now collect together equations (\ref{*Sb1}), (\ref{*Sb2}) and
(\ref{*Sb3}) to deduce that,
\begin{multline} \label{*bdy1}
S^*_\textrm{boundary}=\kappa \int dt \int_{S_t} (-1)^D N^i
\pi_j E^j{}_i 
+N \left\{ 2 \w K^i \wedge \phi_i \right. \\
\left. +4\alpha  \w K^i \wedge
\left[ \w \Omega^{jk} -\w H^j \wedge \w H^k -\frac{1}{3} \w K^j \wedge \w K^k \right] \wedge
\phi_{ijk} \right\}
\end{multline}
\subsection{The conjugate momentum} \label{sec:conjmom}
Now that we have the  action in its correct form,  it remains to calculate the momentum
conjugate to $E^a$. This is given by,
\begin{equation}
p_a=\frac{\partial \mathcal{L}_\textrm{bulk}}{\partial \dot E^a}
\end{equation}
where $\mathcal{L}_\textrm{bulk}$ is the bulk integrand. We could take
$\mathcal{L}_\textrm{bulk}$ from $S^*_\textrm{bulk}$. However, it is
convenient to temporarily undo the integration by parts that gives
equation (\ref{vecNIBP}) (see appendix). In other words, we leave  the derivatives of
$N^a$ in the bulk action. This is perfectly OK, because it does not
affect the bulk dynamics, and therefore the value of the  conjugate
momentum. The bulk integrand is temporarily given by,
\begin{equation}
\mathcal{L}_\textrm{bulk}=\pi_a \wedge N \$_{\bot} E^a -N\mathcal{H}
\end{equation}
where $\pi_a$ and $\mathcal{H}$ are given by
equations (\ref{conjmom}) and (\ref{hamiltonian}) respectively. Using
the zero torsion decomposition (\ref{torsion3}), we can say
\begin{equation}
\mathcal{L}_\textrm{bulk}=-N\pi_a \wedge \left(H^a+l^a{}_b
E^b\right) -N\mathcal{H}=-N\pi_a \wedge H^a-N\mathcal{H}
\end{equation}
where the right hand equation follows from (\ref{symm}), and the
antisymmetry of $l^{ab}$. 

Referring to equations (\ref{torsion3}) and (\ref{timederiv}), we use the chain rule
to show that,
\begin{equation}
p_a=-\frac{1}{N}\frac{\partial \mathcal{L}_\textrm{bulk}}{\partial \w H^a}=\pi_a
\end{equation}
This non-trivial result is due to the following cancelation,
\begin{equation}
\frac{\partial \pi_b}{\partial \w H^a} \wedge \w H^b +\frac{\partial
\mathcal{H}}{\partial \w H^a}=0.
\end{equation}
We conclude that $\pi_a$ is indeed the conjugate momentum. It should
be thought of as a function of $\dot
E^a$. To derive the Hamiltonian, we should invert this
function. However, $\pi_a$ is cubic in $\dot
E^a$, so  the inverse is multivalued. This is a well known property of
higher derivative gravities. In the Hamiltonian picture, this could
mean that we could jump from one solution to another. These
``zigzagging'' histories still provide an extremum of the action. In
this paper, we will assume that at any given time, we have a
unique solution. This is just the same as saying that we are not in
the process of jumping from one solution to another. For more
discussion on multivalued Hamiltonians in this context, refer to \cite{Teitelboim:GBham,Louko:GBham}.
\subsection{The physical Hamiltonian}
We have shown that we can write our action as,
\begin{equation}
S=S^*_\textrm{bulk}+S^*_\textrm{boundary}=\kappa \int dt \left[\int_{\Sigma_t}
\mathcal{L}^*_\textrm{bulk}+\int_{S_t}
\mathcal{L}^*_\textrm{boundary} \right]
\end{equation}
where
\begin{eqnarray}
\mathcal{L}^*_\textrm{bulk} &=& \pi_a \wedge \dot
E^a -N \mathcal{H}-N^a \mathcal{H}_a \\
 \mathcal{L}^*_\textrm{boundary} &=&
N \left\{ 2 \w K^i \wedge \phi_i  +4\alpha  \w K^i \wedge
\left[ \w \Omega^{jk} -\w H^j \wedge \w H^k -\frac{1}{3} \w K^j \wedge \w K^k \right] \wedge
\phi_{ijk} \right\} \nonumber \\
&&\qquad +(-1)^D N^i
\pi_j E^j{}_i .
\end{eqnarray}
The corresponding Hamiltonian is defined as,
\begin{eqnarray}
H &=&\kappa \int_{\Sigma_t} \pi_a \wedge \dot
E^a - \mathcal{L}^*_\textrm{bulk}-\kappa\int_{S_t}
\mathcal{L}^*_\textrm{boundary} \nonumber\\
&=&\kappa \int_{\Sigma_t} N \mathcal{H}+N^a \mathcal{H}_a-\kappa\int_{S_t}
\mathcal{L}^*_\textrm{boundary}
\end{eqnarray}
To arrive at the physical Hamiltonian, we need to subtract off the
background Hamiltonian,
\begin{equation}
\bar H = -\kappa\int_{S_t} 
\mathcal{\bar L}^*_\textrm{boundary}
\end{equation}
Here we have used the fact that the background is a stationary
solution to the field equations~\cite{Hawking:hamiltonian},
\begin{equation}
\mathcal{\bar
H}=\mathcal{\bar H}_a=\bar \pi_a=0.
\end{equation}
The physical Hamiltonian is therefore given by,
\begin{equation} \label{hamphys}
H_\textrm{phys}=\kappa \int_{\Sigma_t} N \mathcal{H}+N^a \mathcal{H}_a-\kappa\int_{S_t}
\Delta \mathcal{L}^*_\textrm{boundary}
\end{equation}
where 
\begin{multline}
\Delta \mathcal{L}^*_\textrm{boundary} = (-1)^D N^i
\pi_j E^j{}_i+ N \left\{ 2  \Delta \w K^i
\wedge \phi_i \right. \\
\left. +4\alpha \left[\Delta \w K^i \wedge
\left( \w \Omega^{jk} -\w H^j \wedge \w H^k\right) -\frac{1}{3} \Delta \left(\w K^i \wedge\w K^j \wedge \w K^k \right) \right] \wedge
\phi_{ijk}\right\} 
\end{multline}
For any quantity $Q$ in the test spacetime with corresponding quantity
$\bar Q$ in the background, $\Delta Q= Q-\bar Q$. Notice that we have
$\Delta \w \Omega^{jk}=\Delta \w H^j=0$. This is because the
geometry of the boundary is the same in the test spacetime, as in the
background.

If our test spacetime is a solution to the field equations, it
satisfies the constraints 
\begin{equation}
\mathcal{H}=\mathcal{H}_a=0
\end{equation}
Its energy is then given by the value of the physical Hamiltonian,
\begin{multline} \label{energy}
E=-\kappa\int_{S_t} (-1)^D N^i
\pi_j E^j{}_i+ N \left\{ 2  \Delta \w K^i
\wedge \phi_i \right. \\
\left. +4\alpha \left[\Delta \w K^i \wedge
\left( \w \Omega^{jk} -\w H^j \wedge \w H^k\right) -\frac{1}{3} \Delta \left(\w K^i \wedge\w K^j \wedge \w K^k \right) \right] \wedge
\phi_{ijk}\right\} 
\end{multline}
Given the technical complexity of Gauss-Bonnet gravity, we believe
that this expression is remarkably simple. Note that for $\alpha=0$,
we recover the correct result for Einstein 
gravity, as of course we should.

\subsection{Using a coordinate basis}
Although the final result (\ref{energy}) is neat and tidy, we might
prefer to work in a coordinate basis, and express the Hamiltonian in
terms of the familiar tensors of General Relativity. In this case, our
canonical variable is the induced metric $\gamma_{ab}$. The conjugate
momentum, $\pi^{ab}$, is given by~\cite{Isenberg:canonical},
\begin{equation}
\pi^{ab} d^{D-1} x=\frac{1}{2} \pi^a \wedge E^b.
\end{equation}
With this in mind, we can verify that the Hamiltonian and momentum
constraints given by equations (\ref{hamiltonian}) and
(\ref{momentum}) respectively, agree with the corresponding
expressions in~\cite{Teitelboim:GBham}. If $z^A$ is the normal to the
timelike boundary, $B$, and $n^A$ the normal to $\Sigma_t$, the Hamiltonian (\ref{hamphys}) can be
written
\begin{equation} 
H_\textrm{phys}=\kappa \int_{\Sigma_t} d^{D-1}x \left[ N \mathcal{H}+N^a \mathcal{H}_a\right]-\kappa\int_{S_t}
 d^{D-2}x ~\Delta \mathcal{L}^*_\textrm{boundary}
\end{equation}
where now
\begin{multline}
\Delta \mathcal{L}^*_\textrm{boundary} = 2N^i\pi_{ij}z^j
+ N \sqrt{\lambda} \left\{ 2  \Delta K \right. \\
\left. +2\alpha (3!)\delta^{[l}_i\delta^{m}_j\delta^{n]}_k\left[\Delta K^i_l 
\left( \w R^{jk}{}_{mn} -2H^j_m H^k_n\right) -\frac{2}{3} \Delta \left(K^i_l K^j_m K^k_n \right) \right] \right\} 
\end{multline}
Here, $\w R^{jk}{}_{mn}$ is the Riemann tensor on $S_t$, constructed
out of the induced metric $\lambda_{ij}$. $K^i_j$ and $H^i_j$ are the
extrinsic curvatures of $S_t$ in $\Sigma_t$ and $B$ respectively, and
$K$ is the trace of $K^i_j$ . More
precisely,
\begin{equation}
K_{ij}=-\lambda_{(i}^l\lambda_{j)}^m \nabla_l z_m, \qquad
H_{ij}=-\lambda_{(i}^l\lambda_{j)}^m \nabla_l n_m, \qquad
K=\lambda^j_i K^i_j.
\end{equation}

\section{Comparison with previous energy expressions}
\subsection{Deser and Tekin}
Now that we have derived an expression for the energy in GB gravity,
we should compare it with previous results in the literature. In
particular, Deser and Tekin~\cite{Deser:quadenergy,Deser:HDenergy}
used a ``conserved charge'' technique to derive the energy of
asymptotically maximally symmetric spacetimes above backgrounds of
constant curvature.  This method can be
applied to generic higher derivative gravities, but we can check it is
consistent with our result in the Gauss-Bonnet case.

Suppose our test spacetime $\mathcal{M}$, is asymptotically
maximally symmetric. We choose our background, $\mathcal{\bar M}$ to be the maximally
symmetric solution with curvature form,
\begin{equation} \label{maxsymm}
\bar \Omega^{AB}=\frac{2 \Lambda_\textrm{eff}}{(D-1)(D-2)}E^A \wedge
E^B
\end{equation}
The effective cosmological constant is given by,
\begin{equation} \label{cosmoconst}
\Lambda_\textrm{eff}=-\frac{1}{4\alpha^*}\left[1 \pm \sqrt{1+8\Lambda
    \alpha^*}\right], \qquad
    \alpha^*=\alpha\frac{(D-3)(D-4)}{(D-1)(D-2)}.
\end{equation}
which is real and multivalued for $1+8\Lambda
    \alpha^* >0$.

Our aim is to calculate the energy associated with the timelike Killing
vector $\partial/\partial t$. We can choose our foliation $\{ \Sigma_t
\}$ so that the shift vector vanishes. Since $\partial/\partial t$ is Killing, it is clear that $\w
H^i=0$. The expression for the energy reduces to,
\begin{equation} 
E=-\kappa\int_{S_t}  N\left\{ 2  \Delta \w K^i
\wedge \phi_i +4\alpha \left[\Delta \w K^i \wedge \w \Omega^{jk} -\frac{1}{3} \Delta \left(\w K^i \wedge\w K^j \wedge \w K^k \right) \right] \wedge
\phi_{ijk}\right\} 
\end{equation} 
Since $S_t$ lies in the asymptotic region, we can say,
\begin{equation}
\Delta \left(\w K^i \wedge\w K^j \wedge \w K^k \right) \wedge \phi_{ijk} \approx 3\Delta
\w K^i \wedge \left( \bar{K}^j \wedge  \bar{K}^k \right)\wedge \phi_{ijk},
\end{equation}
where $\bar{K}^j$ is $\w K^j$ evaluated on the background. The energy
is now given by,
\begin{equation} 
E=-\kappa\int_{S_t}  N\left\{ 2  \Delta \w K^i
\wedge \phi_i +4\alpha \Delta \w K^i \wedge \left(\w \Omega^{jk}-\bar{K}^j \wedge  \bar{K}^k \right)\wedge
\phi_{ijk}\right\} 
\end{equation} 
Using the fact that $\bar H^j=0$, it is clear from equations
(\ref{tony3}) and 
(\ref{maxsymm}) that,
\begin{equation}
\w \Omega^{jk}-\bar{K}^j \wedge  \bar{K}^k =\frac{2 \Lambda_\textrm{eff}}{(D-1)(D-2)}E^j \wedge
E^k.
\end{equation}
This implies that the energy,
\begin{equation}
E=-\kappa\left(1+4 \alpha^* \Lambda_\textrm{eff} \right)\int_{S_t}  2N\Delta \w K^i
\wedge \phi_i=\pm \kappa\sqrt{1+8\Lambda
    \alpha^*}\int_{S_t}  2N\Delta \w K^i
\wedge \phi_i
\end{equation}
where we have used the $S_t$ analogue of the useful formula
(\ref{useful}), and the cosmological constant relation
(\ref{cosmoconst}).

In order to make contact with~\cite{Deser:quadenergy, Deser:HDenergy},
we switch to a coordinate basis,
\begin{equation}
E=\pm \kappa\sqrt{1+8\Lambda
    \alpha^*}\int_{S_t} d^{D-2}x \sqrt{\lambda} 2N\Delta K
\end{equation}
We now follow the procedure described in~\cite{Hawking:hamiltonian}
for Einstein gravity. Let us start with the test spacetime. Near $S_t$, we can express the metric on
$\Sigma_t$ in Gaussian normal coordinates,
\begin{equation}
ds_{\Sigma_t}^2=\gamma_{ab}dx^a dx^b= dz^2+q_{ij}(z, x^k) dx^idx^j.
\end{equation}
where $q_{ij}(0, x^k)=\lambda_{ij}$ is the metric on
$S_t$. Similarly, for
the background, we can write the metric on $\bar \Sigma_t$  (near
$S_t$), as, 
\begin{equation}
d s_{\bar \Sigma_t}^2=\bar \gamma_{ab}d\bar x^a d\bar x^b= d\bar z^2+\bar
q_{ij}(\bar z, \bar x^k) d\bar
x^id\bar x^j.
\end{equation}
To ensure that the normals to $S_t$ agree on the test spacetime and
the background, we choose the diffeomorphism $z=\bar z$ and $x^i=\bar
x^i$.  In these coordinates,
\begin{equation}
K=-\frac{1}{2} q^{ij}q_{ij, z}, \qquad \bar K=-\frac{1}{2} \bar
q^{ij}\bar q_{ij, z}.
\end{equation}
Since both metrics agree on the boundary, we note that $\Delta
q_{ij}=0$ there. Therefore on $S_t$,
\begin{equation} \label{K}
\Delta K=-\frac{1}{2}\lambda^{ij}\left(\Delta
q_{ij}\right)_{,z}=-\frac{1}{2}\left(\Delta q \right)_{,z}.
\end{equation}
where $\Delta q=\lambda^{ij}\Delta
q_{ij}$.  This gives a final energy expression,
\begin{equation} \label{Ecompare}
E=\mp \kappa\sqrt{1+8\Lambda
    \alpha^*}\int_{S_t} d^{D-2}x \sqrt{\lambda}N \left(\Delta q \right)_{,z}.
\end{equation}

We now use Deser and Tekin's method~\cite{Deser:quadenergy,
  Deser:HDenergy} to calculate\footnote{In~\cite{Deser:quadenergy, Deser:HDenergy}, the
  authors do not explicitly write down an energy expression for the GB
  action with a bare cosmological constant $\Lambda$. However, they
  give enough information to easily derive equation (\ref{D+T}).} the energy associated with the timelike
  Killing vector,  $t^A$.
\begin{multline} \label{D+T}
E_\textrm{DT}=\mp \kappa\sqrt{1+8\Lambda
    \alpha^*}\int_{S_t} dS_A n_B\left\{t_C \bar \nabla^B h^{AC}-t_C
    \bar \nabla^A h^{B C}+t^B \bar \nabla^A h -t^A \bar
    \nabla^B h \right. \\
\left. +h^{B C} \bar \nabla^A t_C-h^{AC}\bar \nabla^B
    t_C+t^A \bar \nabla_C h^{B C}-t^B \bar \nabla_C
    h^{AC} +h \bar \nabla^B t^A \right\}
\end{multline}
where $h^{AB}=g^{AB}-\bar g^{AB}$ and $h=\bar g_{AB}h^{AB}$. Here $g_{AB}$
and $\bar g_{AB}$ are the metrics on $\mathcal{M}$ and $\mathcal{\bar
  M}$ respectively. We will choose to work in a synchronous gauge for which
$n^Ah_{AB}=0$~\cite{Hawking:hamiltonian}. As the
metrics agree on the boundary we can also set $h^{AB}=0$ on
$S_t$. If $S_t$ has {\it inward}
pointing normal $\partial /\partial z$, the measure is given by 
$dS_A=-d^{D-2}x \sqrt{\lambda}\delta^z_A$.  For vanishing shift
vector, we have $t^A=N n^A$, and the energy (\ref{D+T})
simplifies to,
\begin{equation}
E_\textrm{DT}=\pm \kappa\sqrt{1+8\Lambda
    \alpha^*}\int_{S_t} d^{D-2}x  \sqrt{\lambda} N \left( \partial_b h^{zb}-\partial^z h
    \right).
\end{equation}
In the Gaussian Normal coordinates
we have recently described, $h^{zb}=0$ and $h=\Delta q$. Deser and Tekin's
energy now reads,
\begin{equation}
E_\textrm{DT}=\mp \kappa\sqrt{1+8\Lambda
    \alpha^*}\int_{S_t} d^{D-2}x \sqrt{\lambda} N\partial^z \left(
\Delta q \right).
\end{equation}
This expression clearly agrees with the equation (\ref{Ecompare}). We can conclude that although our derivation
was very different to that in~\cite{Deser:quadenergy, Deser:HDenergy},
our results are consistent.
\subsubsection{Application to GB black holes} \label{GBblackholes}
One of the nice features of GB gravity (\ref{GBaction1}) is that it
  contains  static, spherically symmetric
  solutions~\cite{Boulware:stringgen,Crisostomo:BHscan,
  Banados:HDBH, Myers:HDBH, Cai:AdSGBBH, Cai:dimconBH, Cho:thermal,
Louko:GBham} of the form,
\begin{equation} \label{GBBH}
ds^2=-V(r)dt^2+\frac{dr^2}{V(r)}+r^2 d\Omega^2_{D-2}
\end{equation}
where $d\Omega^2_{D-2}$ is the metric on a unit $(D-2)$-sphere. We
will assume that $1+8 \Lambda
\alpha^* >0$, so that there are two possible branches for the  potential, 
\begin{equation}
V(r)=1+\frac{r^2}{2(D-1)(D-2)\alpha^*}\left(1 \pm \sqrt{1+8 \Lambda
\alpha^*+\frac{4(D-1)(D-2)\alpha^* \mu}{r^{D-1}}} \right)
\end{equation}
Here $\mu\geq 0$ is a constant of integration that gives mass to the
spacetime. The upper branch has a naked singularity at $r=0$, whereas
the lower branch is a real black hole with a unique event horizon
surrounding the singularity~\cite{Crisostomo:BHscan}. We wish to
calculate the mass, $M$, of these spacetimes above the appropriate
maximally symmetric background, 
\begin{equation}
ds^2=-\bar V(r)dt^2+\frac{dr^2}{\bar V(r)}+r^2 d\Omega_{D-2}^2
\end{equation}
where
\begin{equation}
\bar V(r)=1+\frac{r^2}{2(D-1)(D-2)\alpha^*}\left(1 \pm \sqrt{1+8 \Lambda
\alpha^*} \right).
\end{equation}
The foliation in each case is given by surfaces of constant $t$, so we
soon see that,
\begin{equation}
q_{ij}=r^2(z)\chi_{ij}, \qquad \bar q_{ij}=\bar r^2(z)\chi_{ij}
\end{equation}
where $\chi_{ij}$ is the metric on the unit $(D-2)$-sphere, and
\begin{equation}
\frac{dr}{dz}=-\sqrt{V}, \qquad \frac{d\bar r}{dz}=-\sqrt{\bar V}.
\end{equation}
If $r(0)=\bar r(0)=R$, on $S_t$, it follows that,
\begin{equation}
(\Delta q)_{, z}=-\frac{2(D-2)}{R} \Delta
\sqrt{V(R)} \approx -\frac{(D-2)}{R\sqrt{\bar V}}\Delta V(R) \qquad
\textrm{on $S_t$}.
\end{equation}
For large $R$, 
\begin{equation}
\Delta V(R) \approx \pm \frac{1}{\sqrt{1+8\Lambda
\alpha^*}} \frac{\mu}{R^{D-3}}
\end{equation}
Finally, we note that $N=\sqrt{V}$ and apply equation (\ref{Ecompare})
to give,
\begin{equation}
M=\kappa \Omega_{D-2}(D-2)\mu
\end{equation}
where $\Omega_{D-2}$ is the volume of the unit
$(D-2)$-sphere. This is the
standard result. It is always valid for $\Lambda_\textrm{eff} \leq
0$. For $\Lambda_\textrm{eff}>0$ our analysis is valid only if the de
Sitter horizon is much larger than the black hole horizon~\cite{Deser:HDenergy}.
\subsection{A special case} \label{specialcase}
In the last section, we assumed that $1+8 \Lambda
\alpha^*>0$. Now consider what happens when $1+8 \Lambda
\alpha^*=0$. We cannot make use of the expression (\ref{Ecompare})
because it involves multiplying an infinite integral, by zero! We will
not worry about how one would modify the approach
of~\cite{Deser:quadenergy, Deser:HDenergy} to accomodate
this. Instead, we will sell the approach developed in this paper. Let
us focus on the $5$-dimensional black hole with,
\begin{equation}
V(r) =1+\frac{r^2}{4\alpha} - \sqrt{\frac{\mu}{2 \alpha}}.
\end{equation}
To calculate the mass, we need to go back to equation
(\ref{energy}). We also need to choose a background. In this example,
the maximally symmetric solution with $\mu=0$ is not necessarily the
most natural choice. We might prefer $\mu$ to be chosen so that
the horizon has zero area~\cite{Crisostomo:BHscan}. Whatever our
choice, we illustrate the flexibility of this work by allowing for
non-maximally symmetric backgrounds. We will keeps things general  and say that the background
potential is given by
\begin{equation}
\bar  V(r) =1+\frac{r^2}{4\alpha} - \sqrt{\frac{\bar \mu}{2 \alpha}}.
\end{equation}
As before, our foliation is made up of surfaces of constant $t$, with boundary
 $S_t$ given by $r=R$. To apply the energy expression (\ref{energy}), we need the following
ingredients,
\begin{equation}
N=\sqrt{V},\qquad N^i=0, \qquad \w K^i=\frac{\sqrt{V}}{R}E^i, \qquad \w H^i=0, \qquad
\w \Omega^{jk}=\frac{1}{R^2}E^j \wedge E^k.
\end{equation}
Now use the useful formula (\ref{useful}) in (\ref{energy}), to derive the energy,
\begin{equation} \label{weirdenergy}
E=-\kappa \Omega_3 R^3 \sqrt{V} \left\{ \frac{6}{R} \Delta \left(\sqrt{V}\right)+\frac{24
\alpha}{R^3} \left[ \Delta \left(\sqrt{V}\right)-\frac{1}{3} \Delta
\left(V\sqrt{V}\right) \right] \right\}
\end{equation}
To keep things tidy, we write
\begin{equation}
V=y^2-m, \qquad \bar V=y^2-\bar m 
\end{equation} 
where  
\begin{equation} 
y^2=1+\frac{R^2}{4 \alpha}, \qquad
m=\sqrt{\frac{\mu}{2\alpha}}, \qquad \bar m=\sqrt{\frac{\bar
\mu}{2\alpha}}.
\end{equation}
Now for large $y$,
\begin{eqnarray}
\sqrt{V} &=& y \left[1-\frac{m}{2y^2} +\mathcal{O}\left(\frac{1}{y^4}
\right)\right] \\
\Delta \left( \sqrt{V} \right) &=& -y\left[\frac{\Delta m}{2y^2}+\frac{\Delta
(m^2)}{8y^4}+ \mathcal{O}\left(\frac{1}{y^6} \right)\right] \\
\Delta \left( V \sqrt{V} \right) &=& -3y^3\left[\frac{\Delta m}{2y^2}-\frac{\Delta
(m^2)}{8y^4}+ \mathcal{O}\left(\frac{1}{y^6} \right)\right].
\end{eqnarray}
If we plug this back into (\ref{weirdenergy}) we find,
\begin{equation}
E=3\kappa \Omega_3 \left[2\alpha \Delta
(m^2) \right] +\mathcal{O}\left(\frac{1}{y^2} \right)
\end{equation}
Now we send $R$, or equivalently $y$, to infinity, to derive the black
hole mass,
\begin{equation}
M=3\kappa \Omega_3 \Delta \mu.
\end{equation}
If we had chosen the background to be the black hole of zero size, we
would have $\bar \mu=2\alpha$. Our black hole mass would be given by
$M=3\kappa \Omega_3 (\mu -2\alpha )$, which agrees with the
``minisuperspace'' method employed in~\cite{Crisostomo:BHscan}.

\section{Discussion}
In this paper, we have derived a neat and easy to use expression for
the gravitational energy of a solution in Gauss-Bonnet gravity. This
was done using a Hamiltonian approach, much like the one used by
Hawking and Horowitz~\cite{Hawking:hamiltonian} for Einstein
gravity. Given the technical complexity of the derivation, our final
expression (\ref{energy}) is remarkably simple.

There have been other ways of calculating the energy of certain
Gauss-Bonnet solutions~\cite{Crisostomo:BHscan, Deser:quadenergy,
Deser:HDenergy}. We have shown that our Hamitonian approach yields
results that are consistent with these. Each approach has its advantages and
disadvantages, as we will now discuss.

Consider the ``conserved charge'' method given in~\cite{Deser:quadenergy,
Deser:HDenergy}. The authors identify a conserved current associated
with a  timelike Killing vector. The gravitational energy corresponds to the
``charge'' of this current. This method can be applied to generic
higher derivative gravities, of which Gauss-Bonnet gravity is just a
special case. However, the background spacetimes, or vacua, are always
assumed to be maximally symmetric everywhere. That is not to say that this
method cannot be extended to a more general choice of background. This
should clearly be a topic for future research. It would also be
interesting to know how to apply this method to the special case
discussed in section~\ref{specialcase}. 

Similarly, we should also ask if we can extend our Hamiltonian approach to
more 
general higher derivative gravities. It should be fairly easy to
consider the Lovelock action~\cite{Lovelock:einstein},
\begin{equation}
S=\sum_{n=0}^{\left[\frac{D-1}{2} \right]} \alpha_n S_n,  \qquad S_n=\int_\mathcal{M} \Omega^{A_1 B_1} \wedge
\ldots \wedge \Omega^{A_n B_n} \wedge e_{A_1B_1 \ldots A_nB_n}.
\end{equation}
Here we have a linear combination of the first $\left[\frac{D-1}{2}
\right]$ Euler characters, with the surface terms given by the same
combination of Chern-Simons forms~\cite{Myers:HDsurface}. Life would be more difficult
if we wanted to consider  an arbitrary combination of Riemann
tensors, as the surface terms are generally unknown\footnote{Attempts
have been made to derive surface terms for general higher derivative
gravities using auxiliary fields~\cite{Nojiri:finite,
Nojiri:new-confine, Cvetic:GBBH}. However, in the GB case, these
results do not agree with~\cite{Myers:HDsurface}, so one should
proceed with caution. A deeper understanding of boundary terms versus
boundary conditions can be found in~\cite{Pons:traceK}.}.  

The ``minisuperspace'' method used in~\cite{Crisostomo:BHscan} is
closest in spirit to the Hamiltonian approach. The idea is to consider a static, spherically symmetric ansatz
for the metric, and insert it back into the action. The action becomes
one-dimensional, making it easier to fix the boundary term. When we
turn to the Hamiltonian, and evaluate it  on one of the black hole
solutions given in sections~\ref{GBblackholes} and~\ref{specialcase},
we derive the black hole mass. This method is very  simple and easy to
use, but somewhat limited. It can only be applied when the
one-dimensional ``minisuperspace'' model is valid. This is OK for the
black hole spacetimes discussed in~\cite{Crisostomo:BHscan}, but a
more general approach is  clearly desirable. 

The Hamiltonian approach developed in this paper is the
appropriate generalisation. It can be
applied whatever the symmetries of the solution, and without having to
reduce
the number of dimensions. In particular, we will use it to
investigate the generalisation of braneworld
holography~\cite{Padilla:thesis} for Gauss-Bonnet
gravity~\cite{Padilla:GBholog}.

To sum up, we have derived an expression (\ref{energy}) for the energy
of a solution to Gauss-Bonnet gravity. This can be applied whatever our
choice of background, and whatever the symmetries of our
solution. This should, hopefully, give us a platform to investigate
Gauss-Bonnet gravity more thoroughly.

\medskip
\centerline{\bf Acknowledgements}
\medskip
I would like to thank James Gregory for many interesting discussions
throughout this entire project. I would also like to thank him, and
Simon Ross, for proof reading this article. Thanks also go to Christos
Charmousis for introducing me to Gauss-Bonnet gravity during a very
long viva!  Final thanks go to Steven Gerrard and Michael Owen for
scoring the goals that beat United.
AP was funded by PPARC.
\appendix
\section{Rewriting $S_\textrm{kinetic}$} \label{appkin}
We begin with the expression for $S_\textrm{kinetic}$ given in
equation (\ref{skin}). This contains terms like $\$_\bot
H^a$, which need to be eliminated by integration by parts. Using the
fact that
\begin{equation}
\$_\bot \zeta_{a_1 \ldots a_m}=\$_\bot E^{a_{m+1}} \wedge \zeta_{a_1 \ldots a_{m+1}},
\end{equation}
we find
\begin{eqnarray}
S_\textrm{kinetic} 
&=& \kappa \int_{\mathcal{M}} E^\bot\wedge \biggl\{
\pi_a \wedge \$_\bot E^a +4 \alpha \left[ H^b \wedge \$_\bot \tilde \Omega^{cd}-\tilde
\nabla H^b \wedge \$_\bot \tilde \omega^{cd} \right]\wedge \zeta_{bcd}
\biggr\} \nonumber \\
&& -\kappa \int_{\Sigma_\infty} 2H^b \wedge \zeta_b
+ 4\alpha H^b \wedge \Lambda^{cd} \wedge \zeta_{bcd} \label{kinetic1}
\end{eqnarray}
where $\pi_a$ is given by equation (\ref{conjmom}) and $\Lambda^{cd}$
by equation (\ref{lcd}).

If we note that $\theta^{\bot b}=-H^b$ is the only non-zero
component of $\theta^{AB}$ on $\Sigma_\infty$, we can use equations
(\ref{bdy}) and (\ref{GC2}) to show that,
\begin{equation}
S_\infty = \kappa \int_{\Sigma_\infty} 2H^b \wedge \zeta_b
+ 4\alpha H^b \wedge \Lambda^{cd} \wedge \zeta_{bcd}. \label{infinity}
\end{equation}
This will cancel off the second line in equation (\ref{kinetic1}).

We will soon need the following identities,
\begin{equation}
\tilde \nabla \zeta_{a_1 \cdots a_m} \equiv 0 \equiv \tilde \nabla
\tilde \Omega^{cd}. \label{bianchi}
\end{equation}
The left hand side follows automatically from the zero-torsion
condition (\ref{torsion4}), whereas the right hand side is just the
frame-form version of the Bianchi identity. We now make use of
(\ref{bianchi}), and the relation, 
\begin{equation}
E^\bot \wedge \$_\bot \tilde \Omega^{cd}=dt \wedge \tilde \nabla
\left(N \$_\bot \tilde \omega^{cd} \right). \label{riemann}
\end{equation}
to show  that,
\begin{multline}
\int_{\mathcal{M}} E^\bot\wedge\left[ H^b \wedge \$_\bot \tilde \Omega^{cd}-\tilde
\nabla H^b \wedge \$_\bot \tilde \omega^{cd} \right]\wedge \zeta_{bcd} \\
\shoveleft{\mspace{150mu}=  \int dt \int_{\Sigma_t} \left[ H^b \wedge \tilde \nabla 
\left(N \$_\bot \tilde \omega^{cd} \right)-\tilde
\nabla H^b \wedge N\$_\bot \tilde \omega^{cd} \right]\wedge
\zeta_{bcd}}\\
\shoveleft{\mspace{150mu}= -\int dt \int_{\Sigma_t} \tilde d \left[ NH^b \wedge \$_\bot \tilde
\omega^{cd}\wedge \zeta_{bcd} \right]} \\
= \int dt \int_{S_t} NH^b \wedge \$_\bot \tilde \omega^{cd}\wedge \zeta_{bcd} \mspace{203mu}\label{kinsurf1}
\end{multline}
Note that we have applied Stokes' Theorem in the following way. If $\tilde A$ is a $(D-2)$-form on $\Sigma_t$, then 
\begin{multline} 
\int dt \int_{\Sigma_t} \tilde d\tilde A =\int_{\mathcal{M}}dt \wedge  \tilde d\tilde A=\int_{\mathcal{M}}dt \wedge d\tilde A\\=-\int_{\mathcal{M}} d(dt \wedge \tilde  A)=-\int_B dt \wedge \tilde A=-\int dt \int_{S_t} \tilde A \mspace{12mu}
\end{multline}
where we have used the fact that $d \tilde A= \tilde d\tilde A +E^\bot\{\cdots\}$~\cite{Isenberg:canonical}. We now insert (\ref{infinity}) and (\ref{kinsurf1}) into equation (\ref{kinetic1}) to give,
\begin{equation}
S_\textrm{kinetic}=\kappa \int dt \int_{\Sigma_t} \pi_a \wedge
N\$_\bot E^a +\kappa \int dt \int_{S_t} 4 \alpha NH^b \wedge \$_\bot
\tilde \omega^{cd}\wedge \zeta_{bcd} -S_\infty. \label{kinetic2}
\end{equation}
Notice that  $\$_\bot \tilde
\omega^{cd}$ has been removed from the bulk part of
$S_\textrm{kinetic}$. This is because we have
zero torsion. $\tilde \omega^{cd}$ is not an independant dynamical variable, so any
time derivatives of it should indeed disappear from the bulk.

We are not yet finished with $S_\textrm{kinetic}$. Equation
(\ref{kinetic2}) still contains derivatives of the shift vector. From
equation (\ref{dualbasis}), we deduce that
\begin{equation} \label{timederiv}
\$_\bot E^a=\frac{1}{N} \left( \dot E^a
-\$_{\vec N} E^a \right), \qquad \vec{N}=N^a \frac{\partial}{\partial
x^a}=E^a{}_b N^b X_a.
\end{equation}
Since $\vec{N}$ lives entirely on $\Sigma_t$, ~$\$_{\vec N}$ is just
 the   intrinsic Lie derivative on    $\Sigma_t$~\cite{Isenberg:canonical}. Therefore,
\begin{equation}
\$_{\vec N} E^a = i_{\vec N} (\tilde d E^a) + \tilde d ( i_{\vec N}
E^a)=-N^b i_{\left[ \frac{\partial}{\partial x^b} \right]} (\tilde w^a{}_c)E^c +\tilde \nabla (E^a{}_b N^b).
\end{equation}
where $i_Y A$ is the {\it interior product} of the vector $Y$ and the
$p$-form $A$~\cite{Nakahara:i_X}. After some integration by parts we see that,
\begin{equation} \label{vecNIBP}
\int dt \int_{\Sigma_t} \pi_a \wedge \$_{\vec N} E^a= \int dt
\int_{\Sigma_t} N^a \mathcal{H}_a -\int dt \int_{S_t} (-1)^{D} N^a
\pi_b E^b{}_a,
\end{equation}
where the momentum constraint, $\mathcal{H}_a$,  is given by equation
(\ref{momentum}). In deriving this constraint, we have used the fact that,
\begin{equation}\label{symm}
\pi^b \wedge E^c= \pi^c \wedge E^b.
\end{equation} 
This is not obvious but can be shown using the symmetries of $H_{ab}$ and
the Riemann tensor\footnote{The Riemann tensor on
$\Sigma_t$  is given by $\tilde
R^a{}_{bcd}$ where $\tilde \Omega^a{}_b=\frac{1}{2}\tilde
R^a{}_{bcd} E^c \wedge E^d$.} (on $\Sigma_t$).

By inserting equation (\ref{vecNIBP}) into (\ref{kinetic2}), we arrive at the final
expression (\ref{kinetic3}) for $S_\textrm{kinetic}$.

\bibliographystyle{utphys}

\bibliography{padilla}
 
\end{document}